# Local moment formation and magnetic coupling of Mn guest atoms in Bi$_2$Se$_3$: a low-temperature ferromagnetic resonance study

D. Savchenko[1,4], R. Tarasenko[2], M. Vališka[2], J. Kopeček[1], L. Fekete[1], K. Carva[2], V. Holý[2], G. Springholz[3], V. Sechovský[2], J. Honolka[1]

[1] *Institute of Physics of the Czech Academy of Sciences, Na Slovance 2, 182 21 Prague, Czech Republic*

[2] *CMP, Faculty of Mathematics and Physics Charles University, Ke Karlovu 5, 121 16 Prague, Czech Republic*

[3] *Institute of Semiconductor and Solid State Physics, Johannes Kepler University, Altenbergerstrasse 69, 4040 Linz, Austria*

[4] *National Technical University of Ukraine "Igor Sikorsky Kyiv Polytechnic Institute", Peremohy 37, 03056 Kyiv, Ukraine*

**Abstract**

We compare the magnetic and electronic configuration of single Mn atoms in molecular beam epitaxy (MBE) grown Bi$_2$Se$_3$ thin films, focusing on electron paramagnetic (ferromagnetic) resonance (EPR and FMR, respectively) and superconducting quantum interference device (SQUID) techniques. X-ray diffraction (XRD) and electron backscatter diffraction (EBSD) reveal the expected increase of disorder with increasing concentration of magnetic guest atoms, however, Kikuchi patterns show that disorder consists majorly of μm-scale 60° twin domains in the hexagonal Bi$_2$Se$_3$ structure, which are promoted by the presence of single unclustered Mn impurities.

Ferromagnetism below $T_C \sim (5.4 \pm 0.3)$ K can be well described by critical scaling laws $M(T) \sim (1-T/T_C)^\beta$ with a critical exponent $\beta = (0.34 \pm 0.2)$, suggesting 3D Heisenberg class magnetism instead of e.g. 2D-type coupling between Mn-spins in van der Waals gap sites. From

EPR hyperfine structure data we determine a $Mn^{2+}$ ($d^5$, $S = 5/2$) electronic configuration with a g-factor of 2.002 for $-1/2 \rightarrow +1/2$ transitions.

In addition, from the strong dependence of the low temperature FMR fields and linewidth on the field strength and orientation with respect to the $Bi_2Se_3$ (0001) plane, we derive magnetic anisotropy energies of up to $K_1 = -3720$ erg/cm$^3$ in MBE-grown Mn-doped $Bi_2Se_3$, reflecting the first order magneto-crystalline anisotropy of an in-plane magnetic easy plane in a hexagonal (0001) crystal symmetry. Across the ferromagnetic-paramagnetic transition the FMR intensity is suppressed and resonance fields converge the paramagnetic limit of a $Mn^{2+}$ ($d^5$, $S = 5/2$).

**Introduction**

In the past decade, a new class of materials called topological insulators (TIs) has received tremendous interest, because of their fascinating fundamental physics as well as their potential for future spintronic applications [Hasan, Qi]. The physics of TIs was first studied in two-dimensional (2D) HgTe/HgCdTe quantum well systems, which possess one-dimensional spin-polarized helical edge states [König, Bernevig]. The corresponding class of three-dimensional (3D) TIs was later found in melt-grown chalcogenide-type materials $Bi_{1-x}Sb_x$, $Bi_2Te_3$ and $Bi_2Se_3$ with large spin-orbit coupling.

3D TI systems host 2D topological surface states (TSS), which follow a Dirac-like dispersion relation, thereby forming a Dirac point (DP) singularity. Close to the DP, the spin polarization of the TSS is strictly locked to momentum, supporting spin currents in opposite directions. While spin-polarized currents are usually only achieved in low-dimensional electron systems at high magnetic fields, chalcogenide-based TIs intrinsically develop spin polarization as a consequence of their large spin-orbit coupling and time reversal symmetry. For spintronic applications, time reversal symmetry is especially interesting since 180° backscattering is forbidden, allowing electrons to travel free over longer distances along the surface [Pesin]. However, scattering channels can be opened in the presence of magnetic degrees of freedom, which break the time reversal symmetry. In the latter case, transport should be inhibited and an energy gap is predicted to appear at the DP [Liu, Henk].

The possibility to modulate the topological properties and electron transport by magnetic degrees of freedom attracted considerable interest of material scientists to develop new TI materials and to implement magnetic atoms in 3D TIs. A simple strategy to introduce magnetism to TIs is by deposition of magnetic atoms [Scholz, Honolka, Eelbo, Sessi] or superparamagnetic clusters on the surface [Wray]. An attractive alternative approach is to induce a 3D ferromagnetic phase transition within TI material itself, e.g. by introducing dilute amounts of magnetic dopants either by growth from the melt or by molecular beam epitaxy (MBE). Here, MBE growth techniques on insulating substrates are advantageous in view of future TSS transport studies, since bulk contributions can be minimized in thin TI films. For magnetically doped $Bi_2Te_3$ or $Bi_2Se_3$, theory predicts several possible magnetic-interaction mechanisms for inducing ferromagnetic order, including carrier-mediated RKKY, superexchange, or Van Vleck spin susceptibility [Biswas2010, Abanin2011, Efimkin2014, Choi2012].

The interest in embedding 3d transition metals into bismuth chalcogenide TIs received further boost after superconductivity (SC) was reported for Cu impurities in $Bi_2Se_3$ [Hor2010, Kriener2011]. Presently, respective SC pairing mechanisms in such systems are under vivid discussion (see Ref.[Schneeloch2015] and references therein). Generally, the local moment formation of 3d guest atoms as well as possible interatomic ferro- or antiferromagnetic coupling mechanisms determine the compatibility with a SC phase, all crucially dependent on the structural integration into the lattice, e.g. as interstitial or substitutional defect sites. The discovery of non-magnetic Ni atoms (S = 0) on $Bi_2Te_3$ despite an incomplete d-shell (~$d^9$) [Vondracek2016] underlines the complexity of moment formation, and suggests further potential for 3d-induced SC phases in bismuth chalcogenides apart from Cu.

Experimentally, however, studies are still in a fledgling stage, often with contradictive results both for melt-grown bulk samples and thin films of $V_2VI_3$ type. The magnetic dopants tend to enter the host matrix in various ways, from a substitutional position for Bi to interstitial sites within a quintuple layer (QL) or in the Van-der-Waals (vdW) gap between QLs [J.-M. Zhang, J. Růžička]. In addition, often a coexistence of different magnetic phases has been reported, especially at the surface compared to bulk [J. G. Checkelsky].

Reports on $Bi_2Se_3$ and $Bi_2Te_3$ materials of melt-grown Mn- and Fe-doped $Bi_2Te_3$, ferromagnetic phases were found for Mn and Fe concentrations of several percent [Y. S. Hor, V. A. Kulbachinskii]. For Fe-doped $Bi_2Se_3$ inconsistent results have been reported [Y. H. Choi 2011, Y. Sugama], which may be due to formation of magnetic clusters [Y. L. Chen]. A spin-glass state [J. Choi 2005] but also antiferromagnetism [Y. H. Choi 2012] was observed in Mn-doped bulk $Bi_2Se_3$.

For MBE-grown films, doping with 3d elements lead to various findings, e.g. for Cr doping in $Bi_2Se_3$, both antiferromagnetism [W. Liu 2015] and ferromagnetism [P. P. J. Haazen] was reported, whereas for thin films of Mn-doped $Bi_2Se_3$ ferromagnetic order reported [D. Zhang, L. J. Collins-McIntyre].

Here we present a thorough investigation of the magnetic and structural properties of Mn-doped $Bi_2Se_3$ films grown by MBE on insulating $BaF_2$(111) substrates. The electronic and magnetic ground state of Mn dopants are carefully characterized by temperature EPR measurements, a technique which so far is rarely used to characterize magnetically doped topological insulators [Zimmermann2016, Wolos2016, vonBardeleben2013]. In particular we find that the Mn 5/2 spin states remain predominantly in the single-impurity - non-clustered - regime, and exhibits the (0001) magnetic easy axis plane defined by magneto-crystalline energy terms. In the ferromagnetic regime our FMR data can be fitted assuming a magneto-crystalline uniaxial hard axis along the [0001] direction with anisotropy energies in the range of several thousand erg/cm$^3$. A homogeneous distribution of Mn spins is supported by our finding that the transition to a ferromagnetic phase at $T_C$ ~ (5.4 ± 0.3) K is conform to a 3D Heisenberg class magnetism $M(T) \sim (1-T/T_C)^{\beta}$ with a critical exponent $\beta = (0.34 \pm 0.2)$ derived by Arrott-Noakes plots. This does not endorse 2D dominated magnetic coupling between Mn-spins concentrated e.g. within vdW gap sites, as often suggested in the literature.

**Experimental details**

Thin-films of $Bi_2Se_3$ were grown by MBE on insulating $BaF_2$(111) at temperatures between 360°C and 380°C. Since the (0001) basal plane of the trigonal $Bi_2Se_3$ lattice is nearly lattice-matched to the $BaF_2$ (111) surface, 2D layer growth with good crystalline quality can be obtained as supported by in situ reflection high-energy electron diffraction (RHEED). For details of the growth procedure see Ref. [O. Caha]. Compared to the case of $Bi_2Se_3$ growth on Si (111), the difference in thermal expansion with respect to $BaF_2$ is much smaller than with Si, which further improves substantially the layer quality and the stability of the films. Mn-doped $Bi_2Se_3$ thin-films of thickness $t$ = 300, 500, and 700 nm were grown with additional Mn flux under Se-rich conditions to support substitutional incorporation of Mn on Bi sites rather than as interstitial defects [C.L. Song]. Nominal Mn concentrations $x_{Mn}$ are between 6% and 8% per f.u..

The chemical composition of the samples was checked with a scanning electron microscope Tescan FERA 3 equipped with an energy-dispersive x-ray detector (EDX) EDAX Octane as well as an EBSD camera Digiview IV from the same producer. The structure quality of the samples was checked by high-resolution x-ray diffraction (XRD) using a standard laboratory diffractometer equipped with parabolic multilayer primary optics and a 4-bounce monochromator on the primary side, as well as an analyzer crystal and a point detector on the secondary side.

Atomic Force Microscopy (AFM) measurements were carried out at room temperature on an ambient Bruker instrument (Dimension Icon), using the Peak Force Tapping mode with ScanAsyst Air tips (Bruker; stiffness $k$ = 0.4 N/m; nominal tip radius 2nm). Measured topographies have a resolution of (512 x 512) points. Prior to the scans in the Peak Force mode, scanned areas were measured in the contact mode with a harder tip ($k$ = 5 N/m) to remove remaining Selenium capping.

For EPR studies we use an X-band ($\nu$ ~ 9.4 GHz) Bruker ELEXSYS E580 spectrometer in the temperature range from 30 K to 4.2 K. The EPR experiments were carried out using an ER 4122 SHQE SuperX High-Q cavity with microwave power levels of 0.4743 mW, a modulation frequency of 100 kHz, a modulation amplitude of 100 Oe, and the conversion time of 80 ms.

Reference magnetic properties were furthermore studied in the temperature range 2 K - 300 K using a superconducting quantum interference device (SQUID) magnetometer (MPMS 7XL, Quantum Design) with magnetic fields up to 70 kOe applied parallel and perpendicular to the film plane. Fitting the field dependence of magnetization at 300 K allows determination of the temperature-independent diamagnetic signal of the $BaF_2$ substrate, which was subtracted from the measured data. The temperature dependence of the magnetization $M(T)$ was measured in a magnetic field of 100 Oe. Due to the small values of the magnetization, demagnetization effects were supposed to be negligible in the studied systems.

**Results and discussion**

Properties of $Bi_2Se_3$ and Mn-doped $Bi_2Se_3$ materials, both melt- or MBE-grown, are known to be highly sensitive to changing growth parameters. For a reliable discussion of magnetic results by EPR and SQUID we have characterized the very same samples regarding structural properties using XRD and EBSD techniques.

For $Bi_2Se_3$ we expect a rhombohedral structure with space group $R\bar{3}m$ (D3d5). In the [0001] direction (also denoted as *c*-axis), Se-Bi-Se-Bi-Se QLs are stacked in an fcc-like ABCA-BCAB-CABC fashion, where dashes correspond to vdW gaps between the facing Se atomic planes. Each QL has primarily covalent bonding, and inter-QL bonding along the stacking direction is of weak vdW character.

Figure 1 summarizes results from XRD and shows symmetric $\omega/2\Theta$ scans (000*L* scans) of Mn-doped samples along with a reference sample without Mn. The simulated diffraction curve assuming a perfect $Bi_2Se_3$ single crystal is shown for comparison as a blue line. While pure $Bi_2Se_3$ closely follows the simulated pattern, for Mn doped samples with nominal concentrations $x_{Mn}$ = 6.0% - 6.7% diffuse scattering appears as broad bumps adjacent to the narrow diffraction peaks (000.6), (000.15) and (000.21). Mn concentrations of 7 %, roughly correspond to the Mn solubility limit in $Bi_2Se_3$ [L. J. Collins-McIntyre], so that the broad peaks at the XRD curves are probably caused by diffuse x-ray scattering from spurious Mn-rich precipitates such as MnSe.

Apart from these local defects, extended defects like stacking faults (SF) or twinning domains (TWDs) [Tarakina2014, Madlin2010, Jariwala2013, Li2010, Kriegner2017] occur in $Bi_2Se_3$ epitaxial layers. Such defects formation is facilitated by the weak vdW bonding and - as we will show below – is further promoted by Mn incorporation. The ideal rhombohedral $Bi_2Se_3$ lattice can be described as a periodic …ABCABC… stacking of (0001) atomic planes, while SFs can be described either a …ABCBCABC… sequence (intrinsic SF – the plane A is missing) or …ABCBABC… sequence (extrinsic SF – the plane B is inserted), where the position of the stacking defect is highlighted as underlined characters. TWD defects known to be caused by variations in the substrate height [Tarakina2014] and can be described by two adjacent QLs, which are rotated by 180° about the surface normal leading to a …ABCABCBACBA… sequence. TWDs can occur even on perfectly flat substrates, since e.g. MBE growth always has multiple nucleation points allowing different rotational starting orientations. In fact it was shown that vicinal substrates can suppress the formation of TWDs due to ordered step edge orientations on the surface [Guo2013].

The symmetric 000L diffraction geometry in Fig. 1 is sensitive neither to SFs nor TWDs, since they correspond to a shift of the atomic (0001) planes perpendicular to the diffraction vector. However, by azimuthal scans we have shown that TWDs are indeed present both for pure and Mn-doped $Bi_2Se_3$ samples of comparable kind [Tarasenko2016, Kriegner2017].

In EBSD we are able to directly resolve extended planar domains with lateral resolution of ~50 nm as summarized in Fig. 2. Kikuchi patterns were taken at several energies from 5 kV up to 30 keV. Typical Kikuchi patterns are shown in Fig 2b. EBSD has an information depth of the order of tenths of nanometers and increases with electron energy. The fitting of Kikuchi patterns, recorded at each point of the sample assuming the rhombohedral structure of $Bi_2Se_3$, leads to domain maps shown in Fig. 2d and 2e. The respective analyzed areas of the surface are shown in Fig. 2c as SEM images. In the *c*-axis direction no significant disorder is visible according to an intact layered structure with perfect (0001) orientation (Fig. 2d). Deviations of the c-axis direction are $\leq 1°$ and can be attributed to substrate roughness or fluctuation in layer growth from the EBSD mapping. In the azimuthal direction (Fig. 2e), however, 60° domains along (0-110) and (1-100) are resolved, which are of the order of a few micrometers in lateral size and correspond to TWDs. They are separated by special low-energy coincidence lattice sites (CLS) grain boundaries $\Sigma 3$ and are purely rotational, i.e. only a rotation around c-axis is present (see sketch in Fig. 2). The domain amount grows with the manganese content, which is evident when comparing the EBSD map to the pure $Bi_2Se_3$ reference sample ($x_{Mn} = 0\%$) where the lateral size of single domain areas are drastically enlarged by more than a factor 10. We comment here that (i) abrupt changes in the Kikuchi patterns are visible when crossing the domain boundaries, and (ii) the detected domains are not significantly connected to the confidence parameter of the Kikuchi fits. From this we again infer that within the sampling depth of 30keV measurements only one single domain exists, and furthermore that the domain boundaries are predominantly oriented perpendicular to the film plane. This is also in line with the independence of Kikuchi patterns of the electron energies.

The general increase of disorder with Mn concentration is reflected also by the increase of surface roughness revealed by AFM in Fig. 3. In pure $Bi_2Se_3$, typical pyramidal growth with triangular (111) surface symmetries is visible [Kriegner2017], which becomes increasingly corrugated upon Mn doping (see Fig. 3a and Fig. 3b).

Twinning is also observed as rotated pyramidal hillocks as described e.g. in Ref.[Kriegner2017]. In addition, we want to stress that a careful look at AFM data in Fig. 3c shows the presence of large numbers of screw dislocation defects (SDD)s, one of which is highlighted as a zoom-in. SDDs often appear as centers of the pyramids, which suggests partial screw-dislocation-driven growth instead of layer-by-layer growth. Such a behavior has been observed in $Bi_2Se_3$ synthesis using polyols [Zhuang2014], but also other layered vdW systems like the transition metal dichalcogenides $WS_2$ or $WSe_2$. Screw dislocation growth is often triggered by steps at the substrate interface, from where the helical defect can continue through the whole $Bi_2Se_3$ thickness. Our Mn-doped samples exhibit a roughness of $(1.0 \pm 0.4)$ nm RMS.

While EBSD and XRD give information on structural properties on the scale > 50nm, it is important to have local information on where Mn atoms reside in the host matrix when Mn concentrations are increased. During non-equilibrium MBE growth, we expect Mn to integrate in multiple positions such as substituting Bi ($Mn_{Bi}$) or vdW interstitial ($Mn_{vdW}$) sites (see e.g. Ref.[Figueroa2015]). For $Mn_{vdW}$ we expect two possible sites with octahedral and distorted tetrahedral geometries, respectively. It has been proposed that at low doping concentration Mn is incorporated preferentially as $Mn_{Bi}$ with a solubility limit of approximately 7.5 atomic % [L. J. Collins-McIntyre], whereas at higher Mn concentrations increasing interstitial incorporation will occur [D. Zhang]. In each of these positions Mn will have different local magnetic properties, and also Mn-Mn exchange coupling mechanisms will be position dependent.

We have shown recently that despite the increasing disorder in our films, Mn atoms remain predominantly single impurities in the bulk matrix with a Mn $d^5$ oxidation state. Core-level XPS exhibit a sharp multiplet structure, which can be perfectly described by a weakly hybridized atomic-like threefold $^7P_4$, $^7P_3$, and $^7P_2$ multiplet configuration of Mn $d^5$ [Tarasenko2016, Valiska2016]. It proves that Mn atoms do not significantly form 2D or 3D metallic clusters. Clustering can e.g. happen within the vdW gap, where they could preferentially nucleate at defects like TWD boundaries or SDD lines, but also at surface QL edges. Instead Mn can be described well as single atoms in an octahedral $MnX_6$ environment with negatively charged ligands X, in line with both $Mn_{Bi}$ and $Mn_{vdW}$ geometries with X = Se. The octahedral ligands allow dynamic charge transfer, leading to three characteristic configurations $d^5$, $d^6\underline{L}$, and $d^7\underline{L}^2$, where L denotes a hole in the ligand states. The three configurations are basis states for Hartree-Fock calculations, which also include intra-atomic multiplet coupling. Taguchi et al. have shown that the satellite intensities grow when charge-transfer energies $\Delta = E(d^6\underline{L}) - E(d^5)$ are sufficiently small [Taguchi], allowing electrons to efficiently hop into the Mn 3d shell.

Mn impurities have ferromagnetic properties as proven by standard SQUID techniques shown in Fig. 4 for the very same samples (left to right). The temperature dependence of the magnetization is plotted, when cooled in a small field of 100 Oe oriented in- and out-of-plane of the films. Large values of the magnetization are found for the in-plane orientation of magnetic fields $H \perp c$ at low temperatures, which can be associated with the presence of a hard magnetization direction along the $c$-axis [0001] in agreement with previous studies [Tarasenko2016, D. Zhang]. Inflection points of the $M(T)$ data allows to get an estimation of the Curie temperature, which are in the range $T_C \sim (5.2 \pm 0.2)$ for all samples as listed in Table I. The phase transition can be further analysed by generating Arrott-Noakes plots of a set of magnetisation curves $M^T(H)$ at varying temperature $T$ as shown in Fig. 4d. According to the Arrott-Noakes equation $(H/M)^{1/\gamma} = (T - T_C)/T_1 + (M/M_1)^{1/\beta}$, where $T_1$ and $M_1$ are material parameters [Arrott&Noakes1967]. Plotting $M^{1/\beta}$ versus $(H/M)^{1/\gamma}$ for the correct parameter set $(\gamma, \beta)$ should lead to parallel and straight lines for $M(B)$ at different temperatures close to $T_C$. In Fig. 4e our measured data $M^T(H)$ is plotted according to Arrott-Noakes for adjusted parameters $\gamma = (1.3 \pm 1.0)$ and $\beta = (0.34 \pm 0.2)$, where get best linearity of the data (dashed lines are guides to the eye) and a rough estimation $T_c^{Arrot} = (5.4 \pm 0.3)$. Deviations from a linear behaviour are visible both in the very low-field region below 500 Oe and at highest fields, which can have several reasons. At low fields they can be induced by domain formation. Here it is important to note that using a parameter set $(\gamma = 1, \beta = 0.5)$, predicted for a mean-field magnetic behaviour, significantly worsens the linear behaviour. Instead, our results point towards a 3D Heisenberg universitality class with $(\gamma = 1.396, \beta = 0.369)$ as discussed e.g. for diluted Mn-spins in ferromagnetic semiconductor (Ga, Mn)As, where a similar scaling behaviour like ours is found [Wang2016]. From this we have reason to believe that Mn spins do not couple in a 2D fashion as e.g. expected for $Mn_{vdW}$ − type spins arranged as 2D layers in the vdW gap and with weak couplings to adjacent vdW gaps. This interesting finding suggests to study the critical scaling behaviour further on the basis of a wider varyity of Mn concentrations.

According to Ginzburg-Landau theory, the spontaneous magnetisation $M(T)$ measured along an easy axis direction in zero field follows the critical scaling law $\sim (1-T/T_C)^\beta$ for $T < T_C$, and the susceptibility follows $M/H \sim (T-T_C)^{-\gamma}$ for $T > T_c$. In the experimental data of Fig. 4a–c we added the expected scaling laws for the parameter set $(\gamma = 1.3, \beta = 0.34)$ and $T_C = T_C^{Arrot} = (5.4 \pm 0.3)$ K for comparison. Above $T_C$ the data seems well described, while below $T_C$ the 3D critical scaling law gives a very qualitative agreement but predicts a much more abrupt increase of $M$ below $T_C$. This gradual behaviour is a hint for an inhomogeneous distribution of Mn exchange coupling constants, which should smear out the critical scaling behavior. However, caution is demanded since $M(T)$ is always measured in a weak biasing field of 100 Oe, and thus does not strictly represent the spontaneous magnetisation addressed by theory.

Table I: Overview of sample characteristics derived from SQUID data. For the estimation of the Mn concentration $x_{Mn}^{SQUID}$ a non-compensated S = 5/2 spin state of Mn was assumed.

| sample | $t$ [nm] | Nom. $x_{Mn}$ [%/f.u.] | $M_{sat}$ [emu/cm$^3$] | $x_{Mn}^{SQUID}$ [%/f.u.]*) | $T_C$ [K]**) | $2H_c$ [Oe] |
|---|---|---|---|---|---|---|
| MBE3161 | 300 | 6.4 | (31 ± 2) | (5.7 ± 0.5) | (5.2 ± 0.2) | (30 ± 5) |
| MBE3124 | 500 | 6.0 | (34 ± 4) | (6.2 ± 0.5) | (5.2 ± 0.2) | (20 ± 5) |
| MBE2862 | 700 | 8.0 | (28 ± 3) | (5.1 ± 0.5) | (5.1 ± 0.2) | (20 ± 5) |

*) assuming a Mn moment of 5.9$\mu_B$
**) estimated from inflection points in $M(T)$

The magnetic coupling and in particular the magnetic anisotropy of Mn dopant atoms can be characterized by temperature dependent FMR measurements.
FMR signals of Mn-Bi$_2$Se$_3$/BaF$_2$ films plotted in Fig. 5 are dominated by symmetrical Lorentzian lineshapes in the temperature range T < 20 K. In the sample with 700 nm thickness at $T$ < 40 K a small but sharp feature appears at fields of about 3.4 kOe (see inset on the Fig. 5c). It corresponds to a weak isotropic Mn$^{2+}$ EPR line hyperfine structure (A = 100 Oe) sextet centered at g ~ 2.001 according to the -1/2 → +1/2 transition of paramagnetic Mn spins. The magnetic resonance signal, however, is dominated by coherent contributions of ferromagnetically coupled Mn spins, which – as we will discuss below - are almost fully aligned along the field in the FMR field range.

Decreasing the temperature, the anisotropy of the resonance field position and the intensity of the FMR signal increases. The FMR linewidth was found to increase with increasing film thickness. The typical angular dependence of the FMR field position and linewidth is shown in Fig. 6. Both the FMR field position and FMR linewidth show a distinct dependence on the orientation of the magnetic field with respect to the crystal axes of the specimen. The linewidth shows a minimum when measured in the magnetic hard direction $\varphi_H = 0°$, which is in line with Ref. [vonBardeleben2013] but also with recent results on Mn-doped $Bi_2Te_3$ [Zimmermann2016], although in the latter case the minimum is observed at $\varphi_H = 90°$, since the magneto-crystalline easy axis is rotated by 90°. The observed angular dependence of the linewidth is indicative of a uniform mode FMR spectrum, and thus a homogeneously magnetized sample. Additional broadening can be caused by the local variation of the internal magnetic fields connected with the random incorporation of the Mn atoms over the Bi lattice sites.

We want to note here that we have collected spectra both during increase and decrease of magnetic fields (0 → 8 kOe and 8 kOe → 0, respectively). In the case of the sample with largest thickness of 700 nm, at lowest temperatures $T$ = 4.2 K and in-plane fields, a small hysteretic effect $\Delta H$ = 47 Oe appeared in Mn-$Bi_2Se_3$ FMR lines (see Fig. 5c), which is absent in sample substrate reference lines. In the case of the other samples with lower film thickness no hysteresis could be detected. The hysteresis effect is of unknown origin and was not observed in the FMR literature on Mn-$Bi_2Se_3$ systems before [vonBardeleben2013].

Following Chappert et al. [Chappert1986] the general FMR conditions in a hexagonal uniaxial magneto-crystalline anisotropy energy (MAE) landscape with negligible in-plane anisotropy terms can be derived from a simplified free energy in polar coordinates ($z \perp [0001], y \| [0001]$):

$$F = -MH \cdot \sin\theta \cos(\varphi_H - \varphi) + \frac{1}{2}[4\pi M^2 - (K_1 + 2K_2)] \cdot \sin^2\theta \sin^2\varphi + K_2 \cdot \sin^4\theta \sin^4\varphi \quad (1)$$

Where the first two terms corresponds to the Zeeman energy and contributions from the demagnetization fields $4\pi M$. $K_1$ and $K_2$ are the first and second order anisotropy energies. Eq.(1) describes the energy for angles $\varphi$ and $\varphi_H$ of the vector $M$ and $H$ with respect to the

(0001) plane, respectively. $\theta$ is the polar angle of $M$, while $\theta_H = 90°$ by definition. $\varphi_H = 0°$ thus corresponds to fields in the film plane, $H \perp [0001]$, and $\varphi_H = 90°$ to fields perpendicular to the film, $H \perp (0001)$.

The equilibrium position of the magnetization, [$\varphi_{eq}(\varphi_H)$, $\theta_{eq} \equiv 90°$] is then given by:

$$MH \sin(\varphi_H - \varphi_{eq}) = [4\pi M^2 - 2K_1 - 4K_2)]\sin\varphi_{eq}\cos\varphi_{eq} + 4K_2 \sin^3\varphi_{eq}\cos\varphi \qquad (2)$$

The resonance field position can be obtained by using following relation:

$$\left(\frac{\omega}{\gamma}\right)^2 = \frac{1}{M^2 \sin^2\theta}\left[\frac{\partial^2 F}{\partial \theta^2}\frac{\partial^2 F}{\partial \varphi^2} - \left(\frac{\partial^2 F}{\partial \theta \partial \varphi}\right)^2\right]. \qquad (3)$$

As a result, according to Ref. [Chappert1986]:

$$\left(\frac{\omega}{\gamma}\right)^2 = \left[H\cos(\varphi_H - \varphi_{eq}) + (4\pi M - H_A)\cos 2\varphi_{eq} + H_{A2}\left(3\sin^2\varphi_{eq}\cos^2\varphi_{eq} - \sin^4\varphi_{eq}\right)\right] \times \\ \times \left[H\cos(\varphi_H - \varphi_{eq}) - (4\pi M - H_A)\sin^2\varphi_{eq} - H_{A2}\sin^4\varphi_{eq}\right] \qquad (4)$$

where $\omega = 2\pi\nu$ is the microwave angular frequency; $\gamma = \beta \cdot g$ is the gyromagnetic ratio of the spin system in the sample; $\beta$ is the Bohr magnetron, $g$ the spectroscopic splitting factor $g$, $H$ the resonance field, and $H_{A1} = 2K_1/M$, $H_{A2} = 4K_2/M$ are the first and second-order axial anisotropy fields, correspondingly. We define the sum of resonance fields as $H_A = H_{A1} + H_{A2}$.

Results of fitting of Eq. (4) to the measured resonance fields $H(\varphi_H)$, $0° \leq \varphi_H \leq 90°$, are shown in Fig. 7a for the example of 300 nm films at $T = 6$ K for parameters $H_A = -1.099$ kOe and $H_{A2} = +0.079$ kOe. The negative sign of the uniaxial anisotropy field $H_A$ dominated by $H_{A1}$ indeed indicates that the $c$ axis is the hard axis of magnetization and the easy axis lies in the film plane as confirmed by our SQUID results on the very same samples. Corresponding equilibrium positions $\varphi_{eq}(\varphi_H)$ of the magnetization are plotted in Fig. 7b. It shows the largest slope $\partial \varphi_{eq}/\partial \varphi_H$ at $\varphi_H = 90°$, as expected for an out-of-plane hard axis.

Table II shows an overview of respective fitting parameters $H_A$ and $H_{A2}$ for all investigated films in the temperature range from 4.8 K to 15 K using the above described procedure. From Table II it is seen that uniaxial field values $H_A$ decrease with decreasing temperature and are largely suppressed when heated above $T = 6$ K, as expected for randomly fluctuating Mn spins in a paramagnetic state. In line with this, the resonant fields for $\varphi_H = 0°$ and $\varphi_H = 90°$ move towards the above discussed value 3.4 kOe of a paramagnetic $Mn^{2+}$ 5/2 spin. Nevertheless, even at $T = 10$ K the EPR spectra in Fig. 5 still show weak but broad lines typical for FMR, suggesting that $T_c$ is washed out due to inhomogeneities in Mn-related ferromagnetic properties.

Table II: The parameters deduced from a fit to the experimentally observed angular variation of the resonance magnetic field positions for samples with 300, 500 and 700 nm thickness. The corresponding values $M$ were taken from SQUID data at fields of 4 kOe equivalent to the resonance conditions.

| T [K] | t [nm] | $(4\pi M - H_A)$ [kOe] | $H_A$ [kOe] | $H_{A2}$ [kOe] | $K_1$ [$10^3$ x erg/cm$^3$] | $K_2$ [$10^3$ x erg/cm$^3$] |
|---|---|---|---|---|---|---|
| 4.8 | 300 | 1.963 | -1.842 | 0.269 | -3.72 | 0.43 |
|  | 500 | 1.110 | -0.974 | 0.852 | -5.81 | 1.36 |
|  | 700 | 0.511 | -0.376 | 0.711 | -3.46 | 1.13 |
| 6 | 300 | 1.179 | -1.099 | 0.079 | -3.75 | 0.13 |
|  | 500 | 0.771 | -0.680 | 0.651 | -4.24 | 1.04 |
| 8 | 700 | 0.270 | -0.242 | 0.334 | -1.83 | 0.53 |
| 10 | 300 | 0.203 | -0.190 | 0.095 | -0.90 | 0.15 |
|  | 500 | 0.292 | -0.279 | 0.387 | -2.12 | 0.62 |
| 15 | 700 | 0.167 | -0.160 | -0.160 | 1.40 | -0.26 |

In order to derive quantitative values for the anisotropy energies $K_1$ and $K_2$, the magnetization $M$ has to be extracted from SQUID measurements at the respective temperatures and fields in the range of between 3 - 5 kOe. In the following we will discuss the field-dependent magnetization curves $M(B)$ for out-of-plane ($\varphi_H = 90°$) and in-plane ($\varphi_H = 0°$) geometries in detail.

Figure 8 shows magnetization curves $M(H)$ of the very same samples measured via SQUID at $T = 2$ K. Low-field regions are shown enlarged as insets in Fig. 8 (right column), where coercive fields between 20 Oe and 30 Oe are visible. Independent of the field direction, at $T = 2$ K fields of ~ 10 kOe are sufficient to reach 80% saturation of the magnetization. Above 10 kOe, $M(H)$ shows a slow and s-shaped background, which will contain contributions of a paramagnetic phase, confirming the observed isotropic Mn$^{2+}$ EPR line hyperfine structure. This contribution should be well described by a standard Langevin function $L(x) = \coth(x) - 1/x$ with $x = g\mu\mu_0 B / k_B T$, assuming a Mn moment $5.92\mu_B$ according to a Mn$^{2+}$ 5/2 spin state identified independently from EPR and XPS. Here we want to note that at even $T \sim 4$ K the paramagnetic EPR contribution does not have a significant weight of the total FMR spectra, since at small resonance fields of $\sim$ 3.4 kOe, paramagnetic Mn 5/2 spins moments are only aligned by about 8% along the field direction.

The hard magnetic axis becomes once more evident from the predominantly linear field dependence of $M(H)$ up to about 4 kOe measured for fields along the $c$ axis. Within the easy plane $H \perp c$, the magnetization is saturated to a large degree already in magnetic fields of about 1 kOe. We want to stress that the shape anisotropy should have minor contributions to the hard axis direction. For saturated magnetisation values $M_{sat}$ of a few tens of emu/cm$^3$ listed in Table I, demagnetising fields $4\pi M_{sat}$ are at most of the order of 0.2 kOe only. The hard axis ∥ [0001] thus is of magneto-crystalline origin due to Mn 3d-orbitals feeling the crystal field of the Bi$_2$Se$_3$ host matrix. Aside from the linear field dependence, a minor hysteretic feature is observed in the out-of-plane direction in several samples with different Mn concentrations.

The saturated values $M_{sat}$ allows us moreover to estimate the Mn concentration $x_{Mn}^{SQUID}$ assuming the absence of magnetic compensation effects between 5/2 spins per Mn atom. The values in Table I show that $x_{Mn}^{SQUID}$ is close to the nominal values. A larger deviation towards lower values $x_{Mn}^{SQUID}$ for highest nominal concentrations of 8%, could be a hint for partial compensation via spin-glass or antiferromagnetic phases as previously reported in Mn-doped bulk Bi$_2$Se$_3$ [J. Choi 2005, Y. H. Choi 2012]. MnSe precipitates suggested from our XRD results are good candidates, since MnSe with cubic structure forms antiferromagnetic phases at about $T$ = 120K [Pollard1983].

Using the magnetization values from SQUID measurements at temperatures and fields corresponding to EPR conditions, we can now give an estimation of the anisotropy energies $K_1$ and $K_2$. The temperature dependent results are listed in Table II. For $T < T_c$, negative values $K_1$ of the order of few $10^3$ erg/cm$^3$ dominate the magneto-crystalline anisotropy, while $K_2$ is significantly smaller. $K_1$ values of the same sign and similar magnitude were reported before for Mn-doped Bi$_2$Se$_3$ films [vonBardeleben2013]. When the temperature is raised above $T_c$, the fitted values $K_1$ significantly drop in magnitude. The sign reversal of $K_1$ at highest temperature $T$ = 15 K, has to be taken with caution due to the weak FMR signal strength, which makes the fit much less reliable.

**Conclusions**

The EPR spectra of the Mn-Bi$_2$Se$_3$/BaF$_2$ films of thickness $t$ = 300, 500 and 700 nm at temperatures $T$ = 4 – 20 K are dominated by a broad single FMR signal with a strong temperature and angular dependence of resonance field position, linewidth, and intensity. At larger film thickness we were able to detect a weak isotropic Mn$^{2+}$ EPR line hyperfine structure (A = 100 Oe) sextet centered at g ~ 2.001 corresponding to the -1/2 → +1/2 transition of paramagnetic Mn spins.

The observed angular dependence of the FMR linewidth of all samples can be explained by the random incorporation of the Mn atoms over the Bi lattice sites without significant metallic clustering. Fitting the angular dependence of the FMR field position, the first- and second-order magnetic anisotropy constants in a hexagonal (0001) crystal symmetry can be estimated, which gives anisotropy energies $K_1$ = -3720 erg/cm$^3$ and reflects a uniaxial hard axis along the $c$ axis. SQUID results on the very same samples shows that ferromagnetism below $T_C$ ~ (5.4 ± 0.3) K can be best described by critical scaling $M(T) \sim (1 - T/T_C)^\beta$ with a critical exponent $\beta = (0.34 \pm 0.2)$, suggesting 3D Heisenberg class magnetism instead of e.g. 2D-type coupling between Mn-spins in van der Waals gap sites. Nevertheless, we observe a gradual phase transition both in FMR and SQUID data, which points towards a distribution of magnetic coupling constants between Mn spins. At highest Mn concentrations of the about 8% per f.u. we cannot exclude the onset of a partially compensated magnetic state, for example by antiferromagnetic MnSe precipitates.

**Acknowledgments**

This work was supported by the Czech Science Foundation Grant No. P204/14/30062S as well as the Austrian Science Funds, Grant No. F2504-N17 IRON. Experiments performed in MLTL (http://mltl.eu/) were supported within the program of Czech Research Infrastructures (Project No. LM2011025). This work was supported in part by the MEYS CR FUNBIO CZ.2.16/3.1.00/21568 (purchase), CENAM LO1409 and SAFMAT LM2015088 projects (maintenance). Jan Honolka gratefully acknowledges the Purkyně fellowship program of the Czech Academy of Sciences.

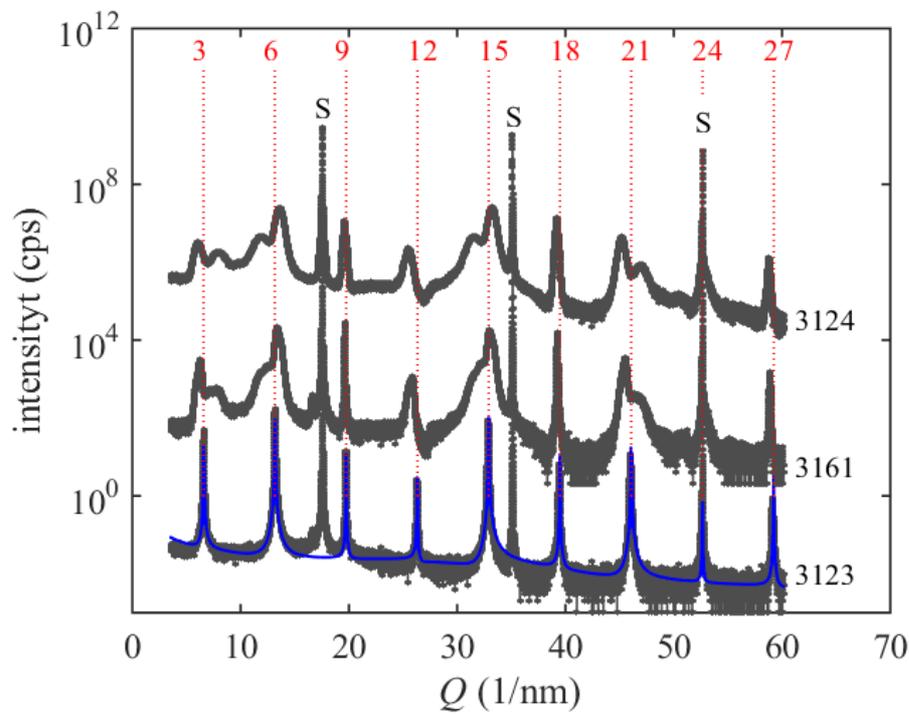

Fig. 1: Measured (points) and simulated (line) symmetric ω/2Θ x-ray diffraction scans along the 000$L$ axis perpendicular to the sample surface with nominal concentrations $x_{Mn}$ = 6.0% and $x_{Mn}$ = 6.4%. The vertical dotted lines denote the theoretical positions of the 000$L$ maxima of the perfect $Bi_2Se_3$ lattice, the corresponding $L$ values are indicated by red numbers (only $L=3n$ maxima are allowed in a symmetric scan). "S" mark the diffraction peaks of the $BaF_2$ substrate. Curves are shifted vertically for clarity, and the parameters of the curves denote the sample numbers. As a reference, the lowest data corresponds to a pure $Bi_2Se_3$ films $x_{Mn}$ = 0%.

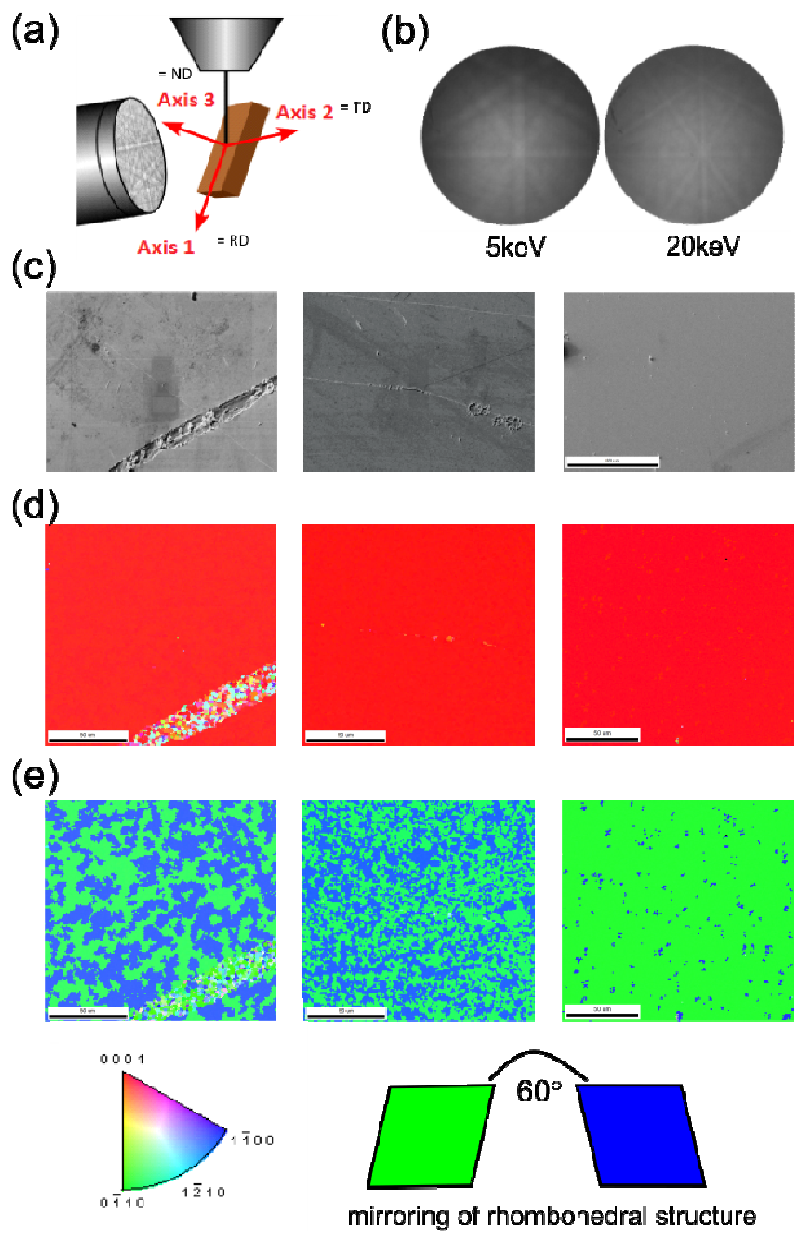

Fig. 2: SEM and EBSD analysis of MBE-grown $Bi_2Se_3$ without and with Mn.
(a) Sketch of the EBSD measurement geometry. (b) Typical Kikuchi patterns of at 5keV and 20keV electron energy. (c) SEM images of the surface morphology for samples with nominal concentrations $x_{Mn}$ = 8.0%, 6.0% and 0% from left to right Mn. (d) and (e) show the respective crystal orientation analysis for polar and azimuthal directions, respectively. According color coding is given on the lower left side.

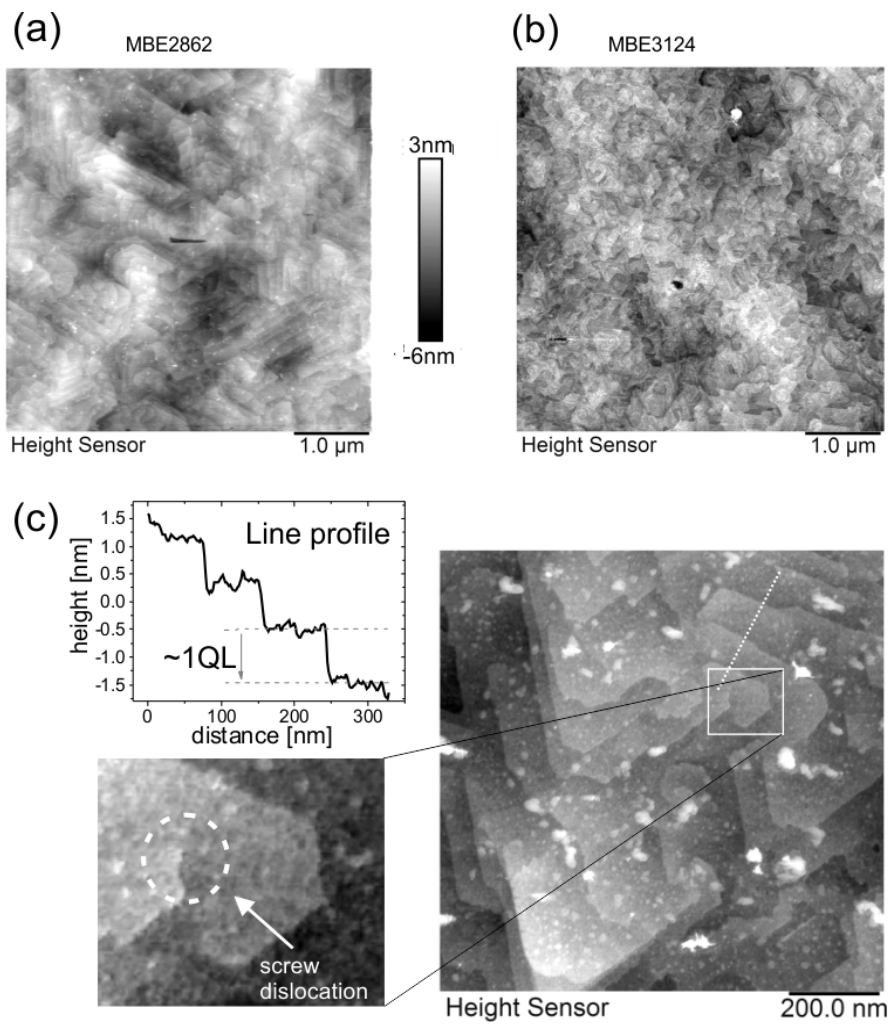

Fig. 3: Atomic force microscopy images (5µm x 5µm) of samples (a) MBE2862 (700 nm) and (b) MBE3124 (500nm) with Mn concentrations $x_{Mn} = 8.0\%$ and $x_{Mn} = 6.0\%$, respectively. (c) (1µm x 1µm) zoom of the MBE2862 surface. A height profile along the dashed line and an example of a typical screw dislocation defect is highlighted as extra images on the left side. The prominent white clusters are reminders of a protective Se cap layer.

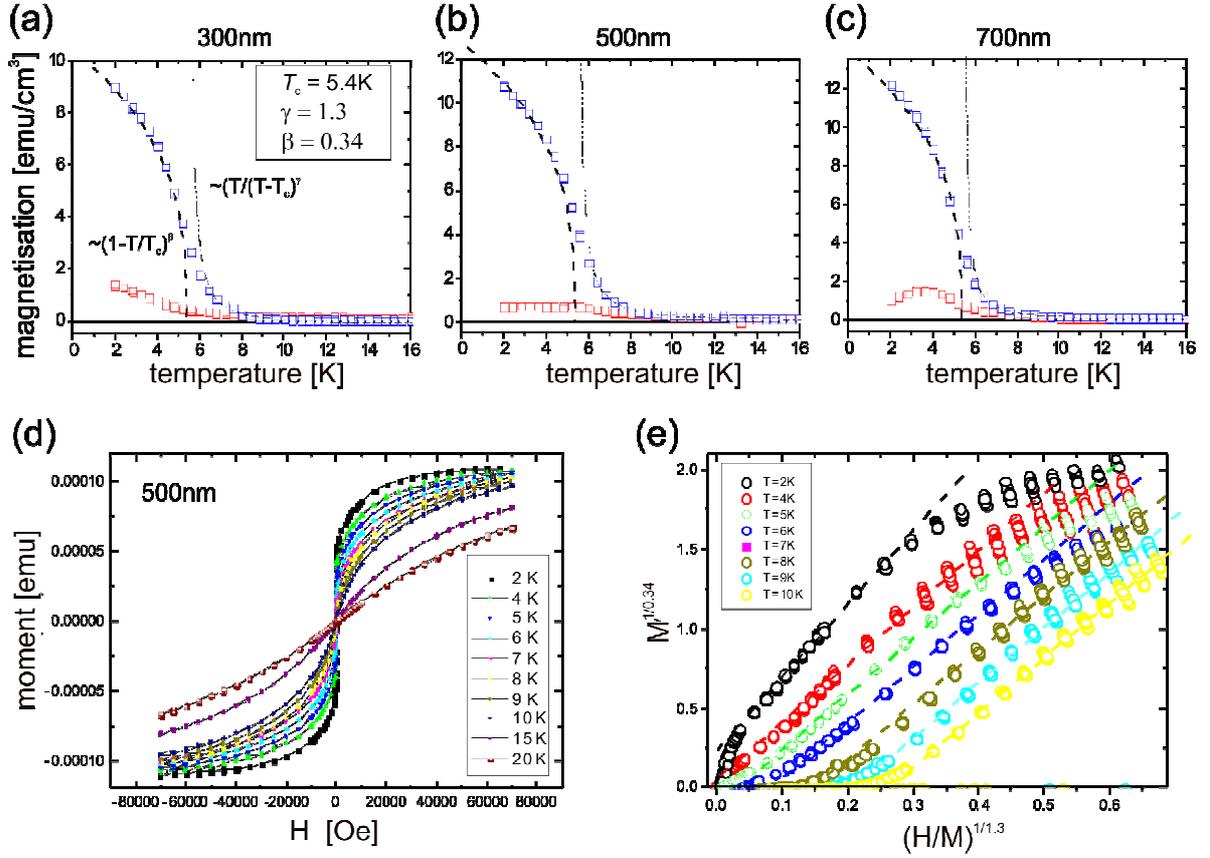

Fig. 4: Temperature and field dependent magnetisation data of Mn-doped $Bi_2Se_3$.
(a), (b), and (c) correspond to the magnetization in units [emu/cm$^3$] of the samples 300, 500, 700 nm and $x_{Mn}$ = 6.7%, 6.0%, and 8%, respectively. The data is measured during cooling in the presence of small bias fields of 100 Oe parallel (red points) and perpendicular (blue points) to the $c$ axis. The dash-dotted lines are fits of critical scaling of the susceptibility at $T > T_c^{Arrot}$ = 5.4 K and γ = 1.3, while the dashed line for $T < T_c^{Arrot}$ corresponds to critical scaling of spontaneous the magnetization $\sim (1 - T/T_C)^\beta$ with the given β derived from the Arrott-plot in (e).

(d) example of temperature dependent in-plane magnetisation curves (sample MBE3124), which in (e) are shown as an Arrott-Noakes plot with γ = 1.3 and β = 0.34 for temperatures around $T_c$. See main text for details.

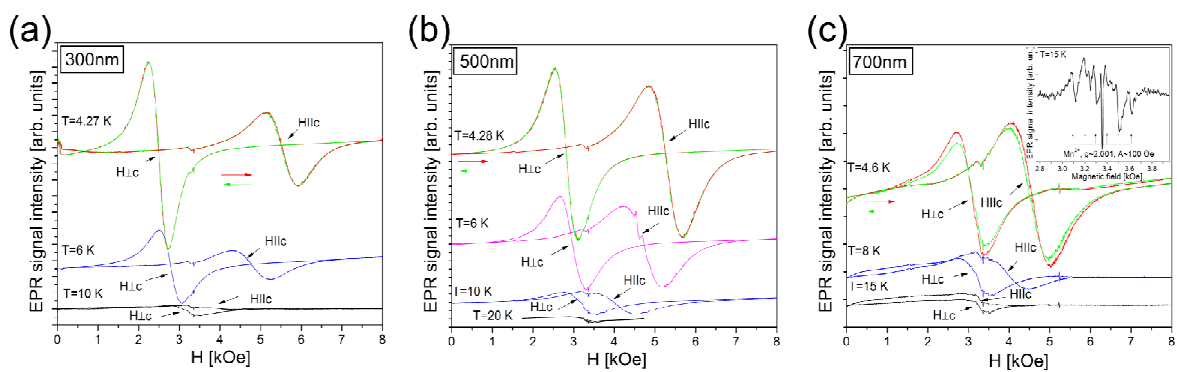

Fig. 5: (a) – (c) EPR spectra collected in both orientations of magnetic field at various temperatures for the films with 300, 500 and 700 nm thickness, respectively. For the case of lowest temperatures $T$ = 4.2 - 4.6 K, EPR spectra measured both in up and down field ramping direction are shown for comparison (see main text for details).

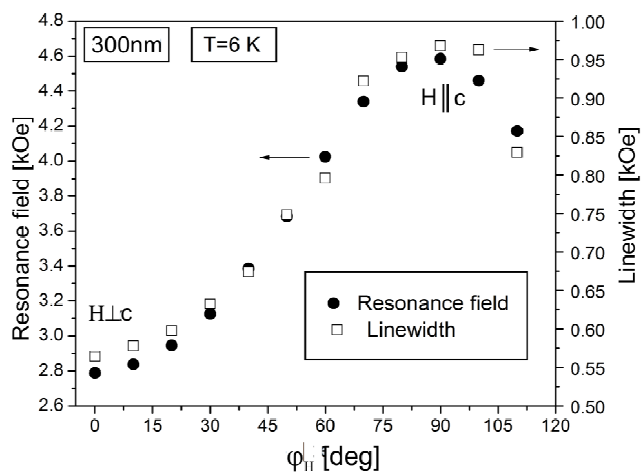

Fig. 6: Angular dependence of the resonance field and linewidth derived from EPR data at $T$ = 6 K. The data is measured on sample MBE3161 with $t$ = 300 nm.

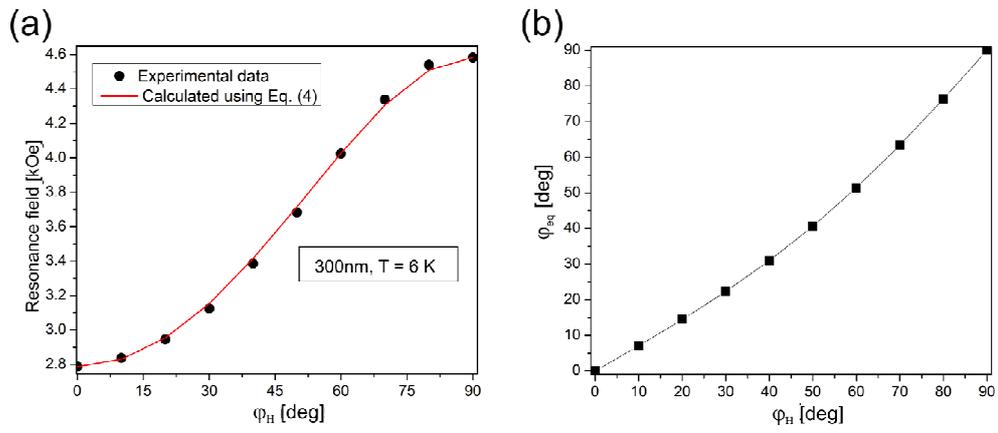

Figs. 7: (a) fitting of the experimental resonance fields using Eq. (4). (b) Equilibrium angle of the magnetization as a function of the magnetic field direction.

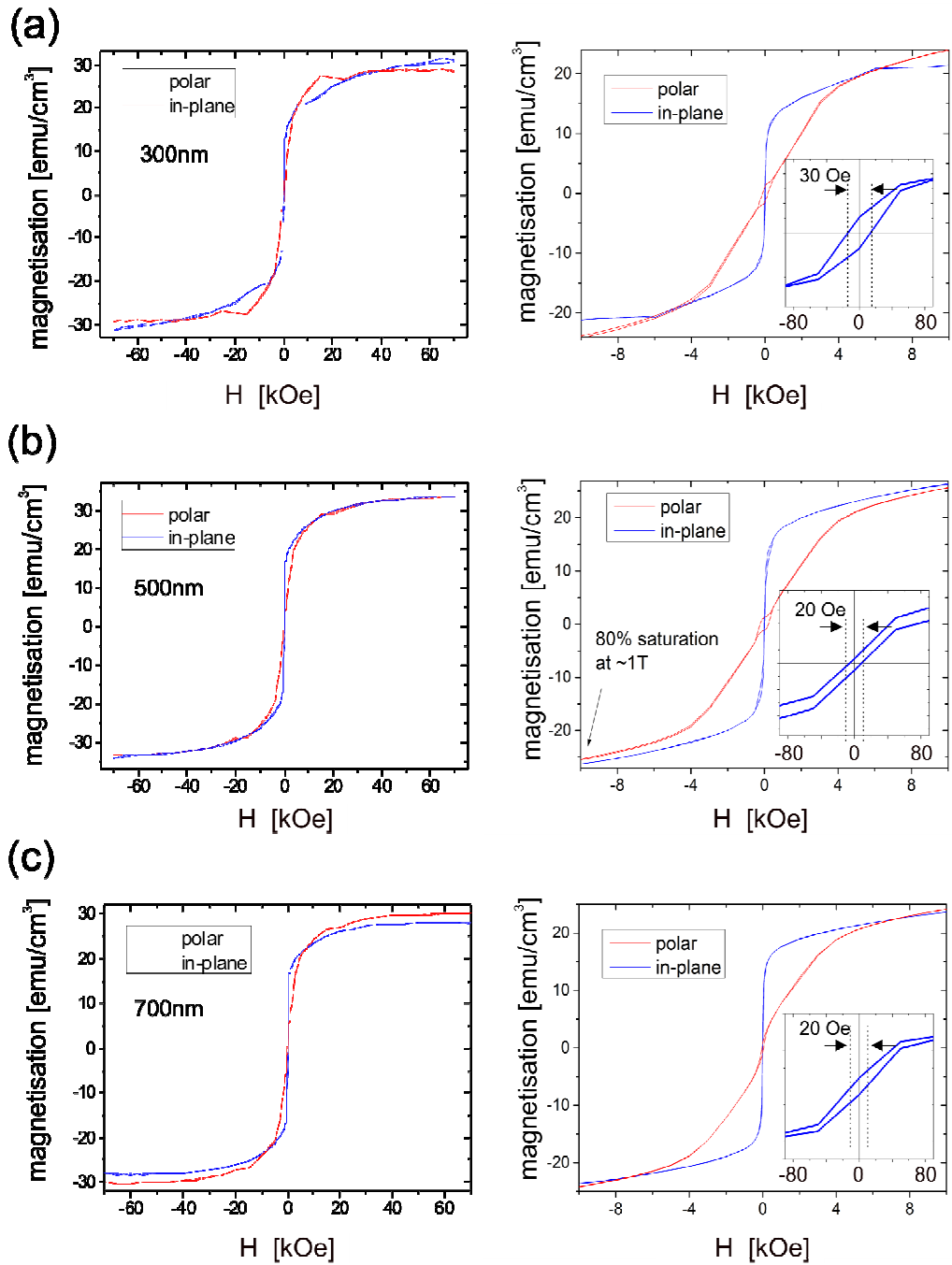

Figs. 8: (a), (b), (c), left column: full field dependence [-70, +70] kOe of the magnetization measured at 2 K of samples 300, 500, and 700 nm and nominal Mn concentrations $x_{Mn}$ = 6.7, 6.0, and 8.0%, respectively. Right column: $M(H)$ at low fields below 10 kOe. The insets show the coercivity in the easy in-plane direction. The field axis is shown in units Oe.